\documentstyle[prl,aps,preprint,tighten,floats,aps,epsf,psfig]{revtex}

\newif\iftightenlines\tightenlinesfalse
\tightenlines\tightenlinestrue

\begin{document}
\def\question#1{{{\marginpar{\tiny \sc #1}}}}
\title{Weighing the universe with accelerators and detectors}
\author{Michal Brhlik\thanks{Electronic mail: {\tt
mbrhlik@umich.edu}}, Daniel J. H.
Chung\thanks{Electronic mail: {\tt
djchung@umich.edu}}, and Gordon L. Kane\thanks{Electronic mail: {\tt
gkane@umich.edu}}}
\address{Randall Physics Laboratory,
        University of Michigan, Ann Arbor, MI 48109-1120}
\date{\today}
\maketitle

\begin{abstract}
Suppose the lightest superpartner (LSP) is observed at colliders, and
WIMPs are
detected in explicit experiments. We point out that one cannot
immediately conclude that cold dark matter (CDM) of the universe has
been observed, and we determine what measurements are necessary before
such a conclusion is meaningful. We discuss the analogous situation for
neutrinos and axions; in the axion case we have not found a way to
conclude axions are the CDM even if axions are detected.
\end{abstract}

\def\question#1{{{\marginpar{\tiny \sc #1}}}}
\def\eqr#1{{Eq.\ (\ref{#1})}}
\def\be{\begin{equation}}
\def\ee{\end{equation}}
\def\mpl{M_{pl}}
\def\sigv{\langle \sigma v \rangle}
\def\mx{m_\chi}
\def\gsim{\lower.7ex\hbox{$\;\stackrel{\textstyle>}{\sim}\;$}}
\def\lsim{\lower.7ex\hbox{$\;\stackrel{\textstyle<}{\sim}\;$}}

\section{Introduction}
Despite the many successes of the standard cosmology, we still do not
know the composition and the amount of energy density in our universe
(for a review, see for example \cite{sadoulet}).  The existence of
matter which does not emit much light is certain from the fact that
stars and other luminous matter contribute only a tiny fraction (about
0.004)\cite{stars} of the critical energy density (required for a
spatially flat universe) while the orbits of stars around galaxies
(see e.g. \cite{begeman}) indicate that the gravitating energy density
is about ten times larger.\footnote{In this paper, we will use the
standard notation of ratio of
energy density in $X$ to the critical energy density as $\Omega_X$.
Since the critical energy density is determined by the Hubble
expansion rate which is uncertain, we will use the usual
parameterization $H=h \mbox{100 km/s/Mpc}$ and often write the energy
density as $\Omega_X h^2$ value instead.}
From a cosmological point of view, a variety of observations, such as
the peculiar motions of galaxies as well as the masses of clusters of
galaxies, corroborate the existence of dark energy and indicate that
$\Omega_M$, the ratio of the average matter-energy density in our
Hubble volume to the critical energy density, is about $1/3$ if the
Hubble expansion rate is about 70 km/s/Mpc as the current observations
indicate (see e.g. \cite{turner}).  Although without other
constraints this dark energy density could be in the form of ordinary
baryonic matter (objects made of neutrons and protons), the standard
model of big bang nucleosynthesis (BBN) calculations and the
measurement of primeval abundance of deuterium gives strong evidence
that the baryon density is $\Omega_B h^2 = (0.02 \pm 0.002)$ which
implies $\Omega_B$ of about $0.04$.  Hence, most of the matter energy
density in the universe seems to be in the form of nonbaryonic dark
matter (NBDM).

Measurements of cosmological parameters will determine $\Omega_{NBDM}$ 
accurately, but are unlikely to help us know the actual character of the
NBDM, since they mainly depend on its gravitational interactions. Some
information on NBDM collisional energy loss and clustering will also
come from astrophysics information. For perhaps the most likely forms
of NBDM we can hope to observe the actual particles in laboratory
experiments, and calculate to a few percent accuracy the actual
contribution to $\Omega$ \cite{kane}. 

Remarkably, the more likely extensions to the standard model of particle
physics provide
candidates for the nonbaryonic dark matter,
candidates which existed even before the need for NBDM was established.  
Indeed, given that these
extensions to the standard model are theoretically compelling from
particle physics considerations, without cosmological considerations,
their provision of dark matter energy of the right order of magnitude
abundance provides an independent hint at the existence of physics
beyond the standard model.  Among the various candidates for this
nonbaryonic dark matter (see e.g. \cite{sadoulet}), a typical unified
supersymmetric theory with conserved R-parity will have a stable
lightest supersymmetric particle (LSP) that will probably constitute
most of the cold dark matter (CDM).  In addition, the non-baryonic
dark energy would most likely consist of neutrino hot dark matter
($\nu$HDM), axionic CDM (ACDM), and the cosmological constant
($\Lambda$).
Given a Lagrangian for the unified theory (UT) and thermal
equilibrium initial conditions for the fluid determining the
cosmological evolution,\footnote{The cosmological initial conditions can 
be determined by an
inflationary model, which in principle can also be determined by the
UT if the initial conditions for inflation are set by an even more
fundamental principle or UT offers a unique inflationary history.
We assume that the underlying cosmology is known when the relic
abundances are finally calculated.}
one can calculate the dark matter content of our universe by solving
the Boltzmann equations.  

Many other NBDM candidates have been proposed \cite{other}. In this
paper we will only analyze the situations for neutrinos, axions, and
LSPs in some detail. Similar conclusions hold for the rest. At
appropriate places in the paper we will consider non-thermal evolution.

The neutrino situation is simpler. Once the cosmology is known and the
neutrino masses are determined, one can compute the relic number of
neutrinos, multiply by the masses, and obtain $\Omega_{\nu}$.
If neutrinos have anomalous interactions they have to be included.

The axion case is difficult. Assume axions are observed in terrestrial
detectors, and perhaps even information about them comes from
astrophysical data. Then the thermal density $\Omega_{a(t)}$ can be
computed precisely enough \cite{kolbturner}. However, coherent oscillations of
axion zero modes give non-thermal contributions $\Omega_{a(nt)}$ that
depend strongly on a ``misalignment'' factor, related to the value of the axion
field at the confinement transition. We are aware of no way to determine 
$\Omega_{a(nt)}$, which can be the dominant contribution, so we conclude
that it may not be possible to even know how much of the CDM is
$\Omega_{axion}=\Omega_{a(t)}+\Omega_{a(nt)}$.

The LSP case turns out to be difficult, but solvable. We first 
demonstrate that knowing the LSP mass, and even knowing its cross section
on a nucleon, do not allow one to know its contribution to 
$\Omega_{LSP}$ to better accuracy than an order of magnitude or so. 
Part of this uncertainty arises from a current lack of knowledge of the 
phases of the soft supersymmetry breaking Lagrangian (${\cal L}_{soft}$),
but even if one arbitrarily took all phases to be zero or $\pi$ at least a 
factor of six uncertainty in $\Omega_{LSP}$ remains. And for an important 
question such as the composition of the CDM, one would not want to make 
unwarranted assumptions about the phases or other relevant parameters.
We then show that knowing the relevant parameters of ${\cal L}_{soft}$,
and $\tan\beta$, to about 5\% allows a determination of  $\Omega_{LSP}$ to
similar accuracy. Such measurements will be possible by combining data 
from hadron colliders, low energy ({\it e.g.} electric dipole moments) 
experiments,  B-factories, and a lepton collider with a polarized beam 
and sufficient energy to produce several superpartners (and appropriate 
luminosity).

If WIMPs are observed in explicit (i.e.,  non-collider) detection
experiments (direct underground 
WIMP scattering on nuclei, or space-based, or indirectly via $\nu$ 
interactions from WIMP annihilation in the sun or earth) there are 
additional large uncertainties which we do not consider in this paper. These 
include going from WIMP-nucleus to WIMP-quark cross sections, local density 
and velocity distributions of WIMPs, how long antiprotons or positrons can 
persist in the galaxy, and so on. All of our results hold even if these
other factors could be controlled.

In the next two sections we discuss the LSP case in detail, first 
analytically and then numerically. Then we turn to some cosmological 
considerations, and examine the axion case. In all cases we assume that
$h^2$ will be known accurately before it is needed to achieve a particle 
physics knowledge of  $\Omega_{NBDM}$, so we do not include errors 
in $h^2$. We also assume that the usual Boltzmann treatment is
sufficient \cite{sred} to estimate the possible uncertainties and that refinements
of the calculational procedure necessary for more accurate calculation
can be accomplished when appropriate. For examples of such refinements,
see Ref.\cite{refine} and references therein.

\section{Analytic estimates}
In this section, we give an analytic estimate for the uncertainties
that can be expected in the calculation of LSP CDM, given that some of
the parameters in the MSSM has been measured to a certain accuracy.
The most common situation for most parts of the MSSM parameter space
is when no other particle mass is within $\mx/20$ of the LSP mass
$\mx$ \cite{coll,wells}.  In that case, self-annihilation determines the relic
abundance.  Assuming as usual that the annihilation products are in
chemical thermal equilibrium with the massless degrees of freedom in a
radiation dominated flat FRW universe, the simplified Boltzmann
equation governing the relic abundance can be written as
\begin{equation}
\frac{d f}{dx} = \mx \sqrt{\frac{ 45 \mpl^2}{4 \pi^3 g_*}} \sigv
(f^2 -f_0^2)
\end{equation}
where $f=n_\chi/T^3$ is the LSP volume density scaled by the cube of
the temperature which sets the scale for the volume density of the
photons, $\sigv$ is the thermal averaged self annihilation cross
section, $x= T/\mx$ is the scaled temperature, $g_*$ counts the
degrees of freedom contributing to the entropy, and
$f_0=\frac{x^{-3/2}}{\sqrt{2 \pi^2}} e^{-1/x}$ is the nonrelativistic
approximation of the thermal equilibrium volume density of LSPs.
Starting from a thermal equilibrium initial conditions, $f$ tracks
$f_0$ until the freeze-out temeperature $x_F$ and then $f$
decouples from $f_0$.  Approximating this decoupling to occur sharply
at a particular temperature $x_F$, one finds an expression for the
relic density as
\begin{equation}
\Omega =T_X^3 \sqrt{\frac{4 \pi^3 g_*}{45 \mpl^2}} (\int_{x_0}^{x_F}
\sigv dx)^{-1}/ \rho_c
\end{equation}
where $\rho_c$ is the critical energy density that only depends on the
cosmology.  The approximate expression for the freezeout temperature
can be written as   
\[
x_{F}\approx \frac{1}{\ln [m_{\chi }\xi \sigv ]+\frac{1}{2}\ln x_{F}}\]
\[
\xi \equiv \frac{1}{(2\pi )^{3}}\sqrt{\frac{45\mpl ^{2}}{2g_{*}}}\]
Propagating the error in quadratures, the fractional error is characterized by
\[
\langle \left (\frac{\Delta \Omega }{\Omega }\right )^{2}\rangle =\sum _{i}(\Delta _{i}(\delta P_{i}))^{2}\]
 where \( \delta P_{i} \) denotes the uncertainty in parameter \( P_{i} \).
Neglecting the \( T \) and \( g_{*} \)uncertainties, we can write 
\[
\Delta _{i}=\frac{[x_{F}^{2}\frac{\partial \mx }{\partial P_{i}}\frac{\sqrt{g_{*}}}{\mx }\sigv +x_{F}^{2}\sqrt{g_{*}}\frac{\partial \sigv }{\partial P_{i}}-\int \frac{\partial \sigv }{\partial P_{i}}\sqrt{g_{*}}dx]}{\int _{x_{0}}^{x_{F}}\sqrt{g_{*}}\sigv dx}\]
 where we have used the fact that \( x_{F}\ll 1 \). We can approximate \( \frac{\partial \sigv }{\partial P_{i}}\approx \sigv \frac{r_{i}}{P_{i}} \)where
\( r_{i}\sim O(1) \) (which is a good approximation in the region of the parameter
space where the \( P_{i} \) dependence is analytic) and conclude
\[
\Delta _{i}\sim \frac{r_{i}}{P_{i}}.\]
 Hence, we conclude that 
\[
\langle \left (\frac{\Delta \Omega }{\Omega }\right )^{2}\rangle =\sum _{i}r_{i}^{2}(\frac{\delta P_{i}}{P_{i}})^{2}\]
 where the strength of the errror contribution is determined by \( r_{i}=\frac{\partial \ln \sigv }{\partial \ln P_{i}} \)which
is what we expect. 

For example, consider a typical nonresonant self-annihilation cross section of
an LSP going to two fermions through a sfermion exchange. We have 
\[
\sigv \approx \frac{g^{2}_{f\tilde{f}\chi }}{64\pi }\sqrt{1-\frac{m_{f}^{2}}{\mx ^{2}}}\left [\frac{c_{1}m_{f}^{2}+c_{2}x\mx ^{2}}{(m_{\tilde{f}}^{2}+\mx ^{2}-m_{f}^{2})^2}\right ]\]
 where \( c_{1} \) and \( c_{2} \) are \( O(1) \) dimensionless constants
and \( g_{f\tilde{f}\chi }^{2} \) is the fermion-sfermion-neutralino coupling.
Then considering the contribution of \( \mx  \) to the fractional error, we
find to leading order in \( \frac{m_{f}}{\mx }\sim \frac{m_{f}}{m_{\tilde{f}}} \)
,
\[
r_{\mx }= \frac{2 \left (1-\left (\frac{m_{\tilde{f}}}{\mx }\right )^{2}\right )}
{1+\left (\frac{m_{\tilde{f}}}{\mx }\right )^{2}}
\]
which is about \( \lsim 1 \) . Considering the contribution of \( m_{\tilde{f}} \)
to the uncertainty, we find similarly
\[
r_{m_{\tilde{f}}}=\frac{-4}{1+\left (\frac{\mx }{m_{\tilde{f}}}\right )^{2}}\]
 which has a magnitude of about \( \gsim 2 \). Finally we see that the coupling
will also contribute
\[
r_{g_{f\tilde{f}\chi }}=2,\] showing that the abundance is most
sensitive to the sfermion mass and the LSP-sfermion-fermion 
coupling. In addition we see that relative uncertainties in parameters are
likely to lead to even larger relative uncertainties in $\Omega_{LSP}$.

The standard scalar neutralino cross section on protons \cite{jung} is
expressed as
\begin{equation}
\sigma_p^{scalar} = \frac{4 m_r^2}{\pi} f_p^2
\end{equation}
where the effective scalar interaction coupling $f_p$ is usually
dominated by the CP-even Higgs parton level exchange between quarks and
neutralinos \cite{drees}
\begin{equation}
f_p^{scalar} \simeq  f_p^{(H)}=m_p \left [\sum_{q=u,d,s} f_{T_q}^p
\frac{f_q^{H}}{m_q}+
\frac{2}{27} f_{TG}^p \sum_{q=c,b,t} \frac{f_q^{H}}{m_q}\right ].
\end{equation}
The matrix element coefficients $ f_{T_q}^p$ and
$f_{TG}^p=1-\sum_{q}f_{T_q}^p$ can be  extracted from pion-nucleon
scattering using chiral perturbation theory and are subject to large
uncertainties which are reflected in the neutralino-proton scattering
cross section \cite{det}.

The parton level couplings, on the other hand, depend entirely on the
SUSY Lagrangian parameters and Higgs masses
\[
f_q^{H}=m_q \sum_{i=1,2}\frac{c_{\chi}^{(i)} c_q^{(i)}}{m^2_{H_i}}
\]
with the (in general complex) neutralino couplings to the light CP-even Higgs
\[
c_{\chi}^{(1)}=\frac{1}{2} (gN^*_{12}-g'N^*_{11})
(N^*_{13}\sin\alpha+N^*_{14}\cos\alpha)
\]
and to the heavy Higgs
\[
c_{\chi}^{(2)}=\frac{1}{2} (gN^*_{12}-g'N^*_{11})
(N^*_{14}\sin\alpha-N^*_{13}\cos\alpha),
\]
where $\alpha$ is the Higgs mixing angle.
The quark-Higgs couplings depend on weak isospin quantum number and we
have
for the up type quarks
\begin{eqnarray}
c_u^{(1)}=-\frac{g}{2m_W}\frac{\cos\alpha}{\sin\beta}, &\ \ \ \ &
c_u^{(2)}=-\frac{g}{2m_W}\frac{\sin\alpha}{\sin\beta}
\end{eqnarray}
and for the down type quarks
\begin{eqnarray}
c_d^{(1)}=\frac{g}{2m_W}\frac{\sin\alpha}{\cos\beta}, &\ \ \ \ &
c_d^{(2)}=-\frac{g}{2m_W}\frac{\cos\alpha}{\cos\beta}.
\end{eqnarray}
In the large $m_A$ limit $\alpha\simeq \beta -\pi/2$ and the $c_d^{(2)}$
coupling is enhanced by a factor of $\tan\beta$. As a result, for small
values of $\tan\beta$ (less than 4) the cross section is dominated by
the light Higgs exchange while for large values of $\tan\beta$ the heavy
Higgs exchange contribution prevails.

In analogy to the analysis of the relic density we can determine
accuracy of the cross section calculation as a function of the variation
in individual parameters
\[
\langle \left (\frac{\Delta \sigma_p }{\sigma_p }\right )^{2}\rangle =
\sum _{i} s_{i}^{2} (\frac{\Delta P_{i}}{P_{i}})^{2}
\]
 where $s_i$ is the variation coeficient corresponding to the uncertainty
\( \delta P_{i} \) in parameter \( P_{i} \).
Assuming that there is indeed a dominant contribution from one of the
Higgs bosons, we find that the Higgs and quark coupling both contribute
to the uncertainty in a similar way
\begin{eqnarray}
s_{c_q}\simeq s_{c_{\chi}}\simeq 2
\end{eqnarray}
and the Higgs mass contribution to the uncertainty is
\begin{eqnarray}
s_{m_H}\simeq -4.
\end{eqnarray}
while the cross section is largely insensitive to variations of the
neutralino mass since the reduced mass of the system is very close to
the mass of the proton.

It is important to realize that the neutralino-Higgs couplings depend
crucially on the neutralino mixing matrix. Both gaugino and Higgsino
components are required to participate if the couplings are not to
vanish. The neutralino mass matrix depends on complex
quantities $M_1$, $M_2$ and $\mu$ with potentially large phases
\cite{bgk} which can significantly modify the lightest neutralino
composition and subsequently its couplings to the Higgs bosons.
It is important to include these complex phases in the general analysis
of the relic density and the neutralino proton cross section in order to
be able to estimate the overall uncertainty in these quantities which
can be achieved once the SUSY Lagrangian parameters including the phases
are measured.




\section*{Numerical Results}

We illustrate the general behavior of the neutralino relic density and
proton elastic cross section on a characteristic set of the MSSM parameters
which can provide neutralino relic abundance in the cosmologically
relevant region and a cross section small enough to be allowed by direct
detection experiments. Since the neutralino scattering cross section
grows with $\tan\beta$ models with low and moderate values of
$\tan\beta$ can easily satisfy direct detection constraints and still
provide a significant neutralino abundance. In all of our numerical
calculations we use the following reference set of parameters --
$M_1=80 {\rm\ GeV}  $,
$m_{A}=250  {\rm\ GeV}$, 
$\tan\beta=3$,
$\varphi_{1}=\varphi_{\mu}=0$, 
$m_{\tilde{\nu}}=110 {\rm\ GeV} $,
$m_{\tilde{\ell}_L}=125 {\rm\ GeV} $,
$m_{\tilde{\ell}_R}=110 {\rm\ GeV} $, 
$m_{\tilde{q}_L}=420 {\rm\ GeV} $,
$m_{\tilde{u}_R}=400 {\rm\ GeV} $,
$m_{\tilde{d}_R}=380   {\rm\ GeV}$.


This set has been chosen so that the resulting neutralino is
predominantly a bino and the effects of the light Higgs and Z
pole neutralino annihilation pole as well as any co-annihilation effects are
minimized. Our choice of parameters leads to the  values of
$\Omega h^2\simeq 0.148 $ and $\sigma_p\simeq 11.4\times 10^{-9} pb$.
Since we are working in a general parametric framework, the soft Higgs masses 
can be chosen so that electroweak symmetry breaking conditions are satisfied.
These numbers are not special, and only illustrate typical results.

\begin{figure}[ht!]
\centering
\epsfxsize=6.25in
\hspace*{0in}
\epsffile{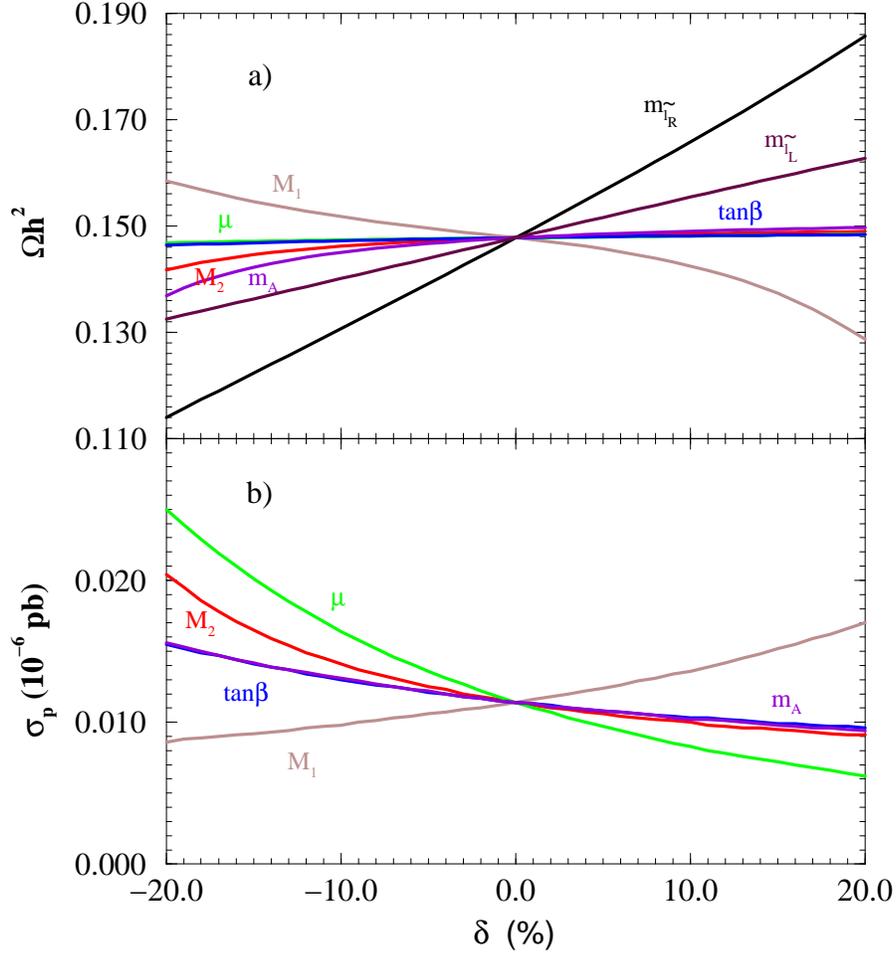}
\bigskip
\caption{Variation of the neutralino relic density {\it (a)} and of the
neutralino-proton elastic scattering cross section  {\it (b)}  with
$\delta=\Delta c_i/c_i$ for the most relevant SUSY parameters.
}
\label{figone}
\end{figure}

First let us turn to the discussion of the neutralino observables
sensitivity to the CP-conserving parameters appearing in the supersymmetric
Lagrangian. Obviously, the most important ones are $M_1$, $M_2$, $\mu$
and  $\tan\beta$ which enter into neutralino mass matrix and determine
both the mass of the lightest neutralino $m_{\chi}$ and its Higgs
and lepton-slepton couplings. Another significant variation comes
from the change in the mass of the CP-odd Higgs boson which influences
the neutralino-Higgs couplings and the heavy CP-even Higgs boson mass.
The dominant slepton exchange contribution to the neutralino
annihilation cross section depends crucially on the slepton mass while
the scattering cross section is insensitive to the scalar masses.

Fig. 1 shows the dependence of the neutralino relic density and proton cross
section on the relative variation of the important parameters within a
$\pm 20\%$ range. As shown in frame {\it a)}, the relic density is
mostly  sensitive to variations in the slepton masses appearing in the 
dominant annihilation diagram -- as the slepton mass increases the 
annihilation cross section is more suppressed and $\Omega h^2$ increases. 
All the other parameters enter the 
neutralino sector and their variations are reflected in the 
variations of the LSP mass and the LSP-slepton-lepton coupling. The most 
significant is the variation of the density with the bino mass   
$M_1$ which determines $m_{\chi}$ and subsequently the dominant $s$-wave
contribution to the neutralino annihilation cross section. It is
relatively stable with respect to changes in $M_2$, $\mu$ and
$\tan\beta$ since the annihilation cross section depends weakly on the
neutralino composition.   In summary, the overall variation
of $\Omega h^2$ is at most $\pm 25\%$ for any single parameter in the 
given range of the input SUSY parameters.

On the other hand, the spin independent part of the neutralino-proton
scattering cross section depends crucially on the gaugino-Higgsino
composition of the lightest neutralino since both components take part
in the neutralino-Higgs interaction. This fact is reflected
in frame {\it b)} which shows a comparable sensitivity to $M_1$, $M_2$
and $\mu$. It is obvious that for the given range of $\delta=\Delta c/c$
the change in $\sigma_p$ can be as big as a factor of two. Note that 
in our particular case ($\tan\beta=3$) the cross section is dominated by 
the light Higgs boson exchange and consequently the sensitivity to $m_A$ 
is limited. As $\tan\beta$ increases, the heavy Higgs exchange takes over 
the cross section and the sensitivity to $m_A$, which determines the heavy 
Higgs boson mass, increases correspondingly. 

\begin{figure}[ht!]
\centering
\epsfxsize=5.in
\hspace*{0in}
\epsffile{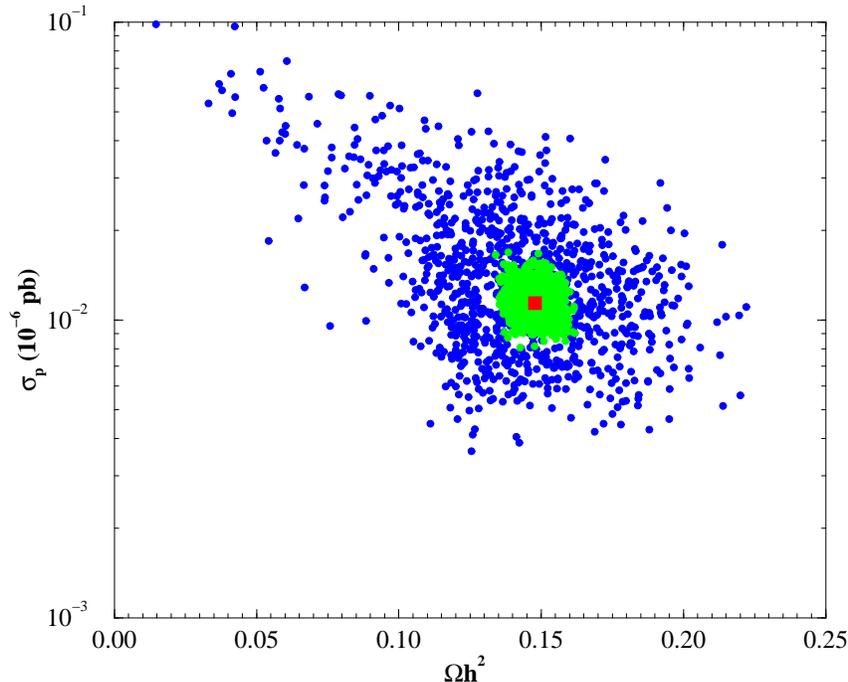}
\bigskip
\caption{Plot of regions in the $\Omega h^2$--$\sigma_p$ plane resulting 
from varying all relevant parameters from Fig. 1  
around their standard value (square) within 5\% (green (grey) region)
and 20\% (dots). The CP-violating phases are set to zero.
}
\label{figtwo}
\end{figure}

In order to estimate the accuracy with which we can determine the relic
density and the proton scattering cross section once we have measured
the SUSY parameters at a collider experiment we plot the range of
both quantities when the SUSY parameters are all varied within a 20 \%
range from the central value. Fig. 2 shows the resulting region in the
$\Omega h^2$--$\sigma_p$ plane. In part of parameter space the variations 
of Fig. 1 can combine, and lead to much larger ranges of variation in 
$\Omega h^2$. In particular, the points in the upper left region with 
small relic density result from non-linear effects associated with 
the presence of an s-channel light Higgs exchange annihilation pole,
which can drastically increase the annihilation cross section, and 
parameter points on the tail of the Breit-Wigner resonance have small
neutralino relic density. This is a fairly generic situation since 
the allowed range for neutralino relic abundance imposes a limit on the 
neutralino mass. Note that the 5\% error region allows a reasonable 
determination of $\Omega h^2$. To understand the weak constraints from 
direct detection experiments, imagine a horizontal line across Fig.2, with 
a height uncertainty coming from the nuclear physics and astrophysics 
ambiguities in extracting $\sigma_p$, and notice the resulting
uncertainties in $\Omega h^2$.

\begin{figure}[ht!]
\centering
\epsfxsize=4.75in
\hspace*{0in}
\epsffile{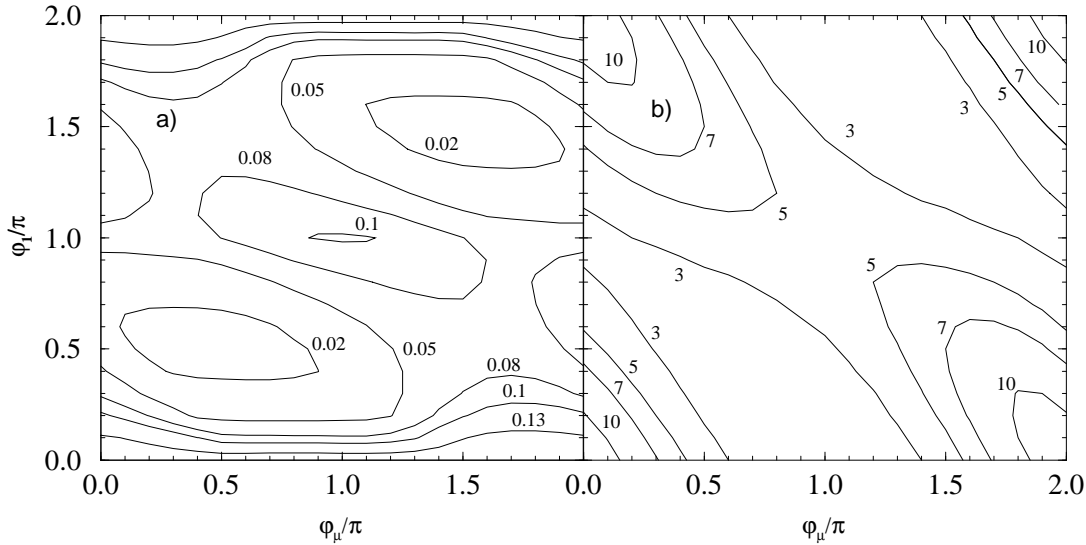}
\bigskip
\caption{Variation of the relic abundance $\Omega h^2$ {\it (a)} and LSP 
proton cross section $\sigma_p$ {\it (b)} (in $10^{-6}\, {\rm pb}$) with 
CP-violating phases affecting the neutralino mass matrix. All other 
parameters are set to the standard set values.  
}
\label{figthree}
\end{figure}

Next we examine the effects of soft SUSY phases on these questions. It has 
been demonstrated recently \cite{bgk} that all soft SUSY phases can be large.
Even if it turned out that some are smaller, {\it e.g.} from future electric
dipole moment experiments, others relevant here may not be. Also, phases have
two effects -- not only do they directly enter the calculations \cite{phase}
of $\Omega h^2$ and $\sigma_p$, they also make measurement of other parameters
such as $\tan\beta$ more difficult, and impossible at LEP or hadron 
colliders \cite{bk}.

We work in a phase parametrization consistent with \cite{bgk} and choose 
$\varphi_2=Arg(M_2)=0$.
In Fig. 3 we show the range of $\Omega h^2$  (frame {\it a)}) and 
$\sigma_p$  frame ({\it b)}) as the two 
relevant phases entering the neutralino mass matrix $\varphi_1$ and 
$\varphi_{\mu}$ are varied in their full range while all other parameters
are kept constant. It turns out that the relic abundance has two local 
minima and  maxima and the range is bigger than an order of magnitude 
between the minimum and maximum. The neutralino proton cross section also varies by more than an order of magnitude but it is monotonous in the 
 $\varphi_{\mu}$- $\varphi_1$ plane. From Fig. 3 it is clear that 
by neglecting the phases one can be missing a crucial part of the 
particle physics information needed to determine the LSP abundance.          

\section*{Cosmological uncertainties and axions}

Although the focus of our paper is examining the uncertainties in our knowledge
of dark matter coming from particle physics, here we will briefly comment on
some of the cosmological uncertainties related to the dark matter
determination.
Specifically, we would like to emphasize a point that can easily be overlooked:
only with a firm determination of the cosmological history can the particle
physics data weigh the universe. To make this point concrete, let us consider
some possible difficulties with the usual approach to the dark matter 
calculation
which we have assumed in our paper. First we will consider the effect of energy
density behaving like a cosmological constant today. Then, we will consider
the situation when the dominant contribution to zero pressure dark
energy is not
a thermal relic. In such ``nonthermal'' cases, particle physics data 
generically
cannot determine the ``matter'' energy density contribution because ``history''
or boundary conditions of the particle physics may not be determined by
a fundamental
principle or dynamics. We will choose axionic dark matter to illustrate this
point. Such uncertainties are relevant to all forms of dark
matter. There can also be extra entropy production which has to be included. 

Let us first consider how the existence of energy density with a
negative pressure
\cite{lambda}(which we will for simplicity call \( \rho _{\Lambda } \) ) 
affects the relic LSP density calculations. First, consider the
Boltzmann equations.
Suppose the spatial curvature is negligible as will be the case after 
inflation. Then, we can start with the same Boltzmann equation
\[
\frac{dn}{dt}+3Hn=-\langle \sigma v\rangle (n^{2}-n_{0}^{2})\]
Since conservation of entropy is still valid, we will have
\[
\frac{df}{dt}=-\langle \sigma v\rangle T^{3}(f^{2}-f_{0}^{2})\]
 where \( f\equiv \frac{n}{T^{3}}. \) In the usual scenario, we assume 
radiation dominance and use
\[
t=\frac{1}{2H}\]
This is still a good approximation because the temperature at which the relic
density froze out is \( O(\textrm{GeV})\gg T_{\textrm{nucleosynthesis}} \)
and we require that \( \rho _{\Lambda } \) not destroy one of the pillars of
standard cosmology and that $\Omega_\Lambda$ be a monotonic function of
time. Hence, the rest
of the formalism follows as usual for the determination of the freezeout 
temperature.
As far as the evolution of the relic density after freezeout is concerned,
there is no change from before again because the nucleosynthesis temperature
is much lower than the freezeout temperature. 

Of course, if we do not require
monotonic variation of \( \rho _{\Lambda } \) then the above relation between
the time and expansion rate may no longer valid and the relic abundance 
calculation
may have to be modified. However, this is unlikely because to preserve
the successes
of BBN, the models of \( \rho _{\Lambda } \) then must be fine tuned such that
its energy is significant at the time of dark matter freezeout while
being negligible
at the time of BBN \cite{finetuning}. In the unlikely event that such tuned
history is that of our universe, then the dark matter energy density implied
by the particle physics measurements may be significantly different from what
we have calculated in this paper.

Axion scenarios offer an elegant solution to the strong CP problem. 
Furthermore,
pseudoscalars resembling axions are generically expected \cite{string} from 
effective theories
arising from compactifications of string theories.\footnote{However, most 
embeddings of axions in string compactification models run into trouble
with scales 
although there may be some ways to circumvent the problem\cite{string}.}
Unlike LSPs, the axions are expected to naturally contribute significantly to
the cosmological energy density even when the thermal relic abundance does not
contribute significantly. Even with the assumption of inflation
and sufficiently low reheating temperature \cite{reh} rendering possible
axionic topological defect related contributions irrelevant, one must account
for the fact that generically there may be coherent oscillations of the axion
zero modes (commonly called misalignment contribution\cite{kolbturner}) giving
a nonthermal contribution to the cosmological energy density with the magnitude
\[
\Omega _{a(nt)}h^{2}=0.13\times 10^{\pm 0.4}(\frac{\Lambda_{QCD}}{200\, {\rm  MeV} })^{-0.7}f(\theta
_{1})
\theta _{1}^{2}(\frac{m_{a}}{10^{-5}\,{\rm eV}})^{-1.18}\]
 where \( m_{a} \) is the axion mass, \( f \) is some known monotonically
increasing function accounting for anharmonic effects (\( f(0)=1) \),
and \( 
\theta _{1} \)
is the ``amplitude'' of the oscillations of the axion field. The value of
\( \theta _{1} \) is essentially equal to the value of the axion field at the
confinement transition. Because there is no direct way to measure \( 
\theta _{1}\ \),
even with a precise determination of \( m_{a} \), there will be a great 
uncertainty
in the axionic contribution. This is in contrast with the most likely smaller
\cite{kolbturner}
thermal relic contribution of the axion which can be quite precisely determined
once the mass of the axion is measured. Indeed, most of the relic axion 
detection
experiments such as electromagnetic spectrum observation of nearby clusters,
as well as Sikivie-type resonant microwave cavities immersed in strong magnetic
fields, mostly measure the axion mass, and there is no obvious direct handle
on the ``misalignment'' angle \( \theta _{1} \). Indeed, \( \theta _{1} \)
may be most likely determined randomly from the space of all possible values,
since above the confinement transition the axions have essentially zero mass.
Hence, in such ``nonthermal'' cosmological boundary condition dependent 
situations, the determination of the dark matter from particle physics data
seems impossible.

\section*{Conclusion}

Understanding the composition of our universe is one of the most fundamental
issues we can study. We demonstrate that we can only be fully confident
we have learned the
answer if we can calculate from first principles and laboratory data for each
type of matter X what $\Omega_X$ is, and find agreement with cosmological 
measurement of  $\Omega_{matter}$. For neutrinos this can be done if their
masses can be measured, or shown to be small compared to $\sim 1\,{\rm eV}$.
For axions we have emphasized there does not seem to be a way to do this.

For the LSP it is possible to calculate  $\Omega_{LSP}$ if measurements of
soft supersymmetry phases, and $\tan\beta$ are made. Unless such 
measurements are done, it is simply incorrect to suggest that detection 
of the LSP at a collider, or in a direct or indirect WIMP experiment, means 
that the cold dark matter has been observed. Indeed, there is a loose 
correlation of easier detection with smaller relic density.
All measurements, of course, help constrain the situation, and the
detection of the LSP will mean
that {\it some} of the CDM has been observed -- but perhaps only a fraction.


\end{document}